\def\beq{\begin{equation}} 
\def\eeq{\end{equation}} 
\def\bed{\begin{displaymath}} 
\def\eed{\end{displaymath}} 
\def\beqq{\begin{eqnarray}} 
\def\eeqq{\end{eqnarray}} 
\def\bedd{\begin{eqnarray*}} 
\def\eedd{\end{eqnarray*}}

\def\n{\nonumber}
 
\def\bbbz{{\bf Z}} 
 
\def\bbb1{{\rm 1\!1}}

\newcommand{\refs}[1]{(\ref{#1})}

\documentstyle[12pt]{article} 
 
\advance\hoffset by -.35in 
\begin{document} 
\bibliographystyle{unsrt} 
 
\def\pr{\prime} 
\def\pa{\partial} 
\def\es{\!=\!} 
\def\ha{{1\over 2}} 
\def\>{\rangle} 
\def\<{\langle} 
\def\mtx#1{\quad\hbox{{#1}}\quad} 
\def\pan{\par\noindent} 
\def\lam{\lambda} 
\def\La{\Lambda} 
 
\def\A{{\cal A}} 
\def\dal{\tilde\alpha}
\def\de{\delta}
\def\be{\beta}
\def\ga{\gamma}
\def\G{\Gamma} 
\def\Ga{\Gamma} 
\def\F{{\cal F}} 
\def\J{{\cal J}} 
\def\M{{\cal M}} 
\def\R{{\cal R}} 
\def\W{{\cal W}} 
\def\tr{\hbox{tr}} 
\def\al{\alpha} 
\def\d{\hbox{d}} 
\def\De{\Delta} 
\def\L{{\cal L}} 
\def\H{{\cal H}} 
\def\Tr{\hbox{Tr}} 
\def\I{\hbox{Im}} 
\def\e{\epsilon}
\def\R{\hbox{Re}} 
\def\h{\bar h}
\def\di{\partial\!\!\!\!\,/}
\def\ti{\int\d^2\theta} 
\def\bti{\int\d^2\bar\theta} 
\def\ttbi{\int\d^2\theta\d^2\bar\theta} 
 \def\bD{\bar D}
\def\bh{\bar h}
\def\la{\lambda}
\def\bp{\bar\phi}
\def\W{{\cal{W}}}
\def\La{\Lambda} 
\def\Ga{\Gamma} 
\def\ga{\gamma}
\def\dga{\tilde\gamma}
\def\be{\beta}
\def\dbe{\tilde\beta}
\def\W{{\cal W}} 
\def\L{{\cal L}}
\def\tr{\hbox{tr}} 
\def\al{\alpha} 
\def\dal{\tilde\alpha}
\def\De{\Delta} 
\def\de{\delta}
\def\L{{\cal L}} 
\def\nn{\nonumber}
\def\la{\lambda}
\def\cov{{\bf D}}
\def\Om{\Omega}
\def\Di{\cov\!\!\!\!/}

\begin{titlepage}
\hfill{DTP/99/51}

\hfill{KCL-MTH-99-31}
\vspace*{2cm}
\begin{center}
{\Large\bf Electric Black Holes in Type 0 String Theory}\\
\vspace*{2cm}
Ivo Sachs\\
\vspace*{.5cm}
{\em Department of Mathematical Sciences\\
University of Durham\\
Science Site, Durham DH1 3LE, UK\\
and\\
Department of Mathematics, King's College London\\
The Strand, London WC2R 2LS, UK}\\
\vspace*{2cm}
\end{center}
\begin{abstract}    
We discuss $AdS_{2+1}$ (BTZ) black holes arising in type $0$ string 
theory corresponding to D1-D5 and F1-NS5 bound states. In particular we 
describe a new family of non-dilatonic solutions with only $Dp_{+}$, 
that is ``electric'' branes. These solutions are distinguished by the 
absence of fermions in the world volume theory which is an interacting CFT.
They can not be obtained as a 
projection of 
a type II BPS-configuration. As for previous type 0 backgrounds 
linear stability is guaranteed only if the curvature is of the order of 
the string scale where $\al'$ corrections cannot be excluded. 
Some problems concerning the counting of states are discussed.

\end{abstract}    
   
\end{titlepage}
\section{Introduction} 
String theory has lead to important new insights in the 
microscopic properties of those black holes for which the near horizon 
geometry consists of an Anti-de Sitter ($AdS$) throat times a sphere which 
represents the horizon \cite{Vafa}. In the case of $AdS_{2+1}$ the entropy 
and Hawking radiation can be exactly reproduced within the world volume 
theory of the corresponding brane configuration of type II string 
theory which in this case is a $1+1$-dimensional conformal field theory 
(CFT). In fact, many of the properties of this CFT, in particular, 
the central charge can be obtained solely by analysing the 
generators of the asymptotic symmetries of the near horizon geometry 
\cite{Brown} (see \cite{strominger} for implications for the black hole 
entropy). 
These properties are therefore universal and essentially independent of 
the underlying microscopic theory. In fact $AdS_{2+1}$ gravity itself 
without any extra fields provides the simplest representation of the 
conformal algebra on the boundary \cite{Brown,Coussaert}, 
but there are indications that 
gravitational degrees of freedom alone cannot account for for all 
micro states needed to produce the geometric entropy of black holes 
\cite{Carlip}. 
Another representation of the algebra is, of course, provided 
by the superconformal world volume theory of the $D1D5$-brane 
system of  type II theory \cite{Vafa,SW}. So far this is essentially 
the only realisation for which 
it has been possible to count the micro states explicitly. In that 
sense string theory passes a non-trivial test for a consistent microscopic 
theory of gravity. One of the puzzles raised by this is the role 
of supersymmetry, or the necessity of fermions in any consistent 
theory of gravity. It is therefore of interest to analyse black 
holes in theories without space-time supersymmetry from this point of 
view. As mentioned above pure gravity provides one such theory 
but does not seem to be able to account for all degrees of freedom. 

In this paper we consider type $0$ string theory which is 
distinguished by being supersymmetric on the world sheet but has no 
space-time fermions in the perturbative closed-string sector. As we 
shall see under suitable assumptions about higher order corrections 
there are black hole solutions that 
are indistinguishable from their type II cousins. The world volume 
theory, however, is rather different due to the absence of fermions. 
The main focus of this paper is to find consistent solutions to 
the type $0$ low energy gravitational effective action and to identify 
the corresponding world volume theory. Previous work on the 
$AdS_{p+1}/CFT$ correspondence in type $0$ theories includes 
\cite{Tseytlin,Costa}. In particular we use the results 
of \cite{Tseytlin} where the low energy effective action has been determined 
up to second order in the tachyon. In \cite{Costa} the 
$D1_{\pm}D5_{\pm}$ black hole solution was given and the entropy was 
reproduced within the corresponding world volume theory which is, in 
fact, a $\bbbz_{2}$ projection of the type IIB theory. This solution 
was further analysed in \cite{Myung}. In this paper we shall describe 
a new solution which consists of only $Dp_{+}$ (i.e. 'electric') branes. 
The fact that such a configuration leads to a non-dilatonic (that is 
conformal) background may come as a surprise. Indeed generically 
both electric- and magnetic RR-fields need to be turned on in order to 
cancel the tachyon 
tadpole. However, as we shall see, for the $D1D5$-brane system this 
is not so. 
Another important feature that distinguishes the $D1_{+}D5_{+}$ black 
hole constructed here is that it cannot be obtained by a projection 
of a corresponding type II solution. In particular the world volume 
theory is purely bosonic. This is interesting, in particular in view 
of the the above remarks about the role of supersymmetry in black hole 
physics. On the other hand, as expected, the absence of a close 
relation  
with supersymmetric black holes considerably complicates the analysis 
of the world volume theory in particular. So, for example is is 
unclear at present how the entropy can be reproduced within 
world volume theory. Another important issue concerns the stability of 
the the background. As for all type $0$ solutions constructed so far 
linear stability is guaranteed only as for $AdS$-radii of the order 
of the string scale. Hence $\al'$ corrections to the background 
cannot be excluded on general grounds. However, it has been suggested in  
(e.g.\cite{Tseytlin,Costa}) that $AdS$ backgrounds are 
indeed protected due to the vanishing of the Weyl tensor, but 
further work is clearly needed to decide on the stability. We 
address some of the issues about the world volume theory which in the 
infra red is a $1+1$ dimensional CFT. A more 
in depth analysis, needed for an explicit understanding of the 
counting of micro states will be referred to future work. 

\section{Type $0$ Strings} 
Type $0$ string theory \cite{Dixon} has been discussed in numerous papers. 
For the analysis presented here \cite{Gaberdiel} and 
\cite{Tseytlin} will be most relevant. These papers provide excellent sources 
for the material used here so that we can limit ourselves to summarise the 
main points. 
\vspace{.3cm}
\pan
{\it Spectrum :}\pan
\pan
Here we will be concerned with type $0B$ string theory which is based on a 
non-supersymmetric 
diagonal GSO projection ($\ha[1+(-1)^{F+\bar F}]$), \cite{Dixon}
\beq
(NS{-},NS{-})\oplus (NS{+},NS{+})\oplus(R+,R+)\oplus(R-,R-)
\eeq
which projects out the fermions in the bulk. Alternatively it can be 
obtained by projecting type IIB by the discrete symmetry generated 
by 
$(-1)^{F_{s} } $, where $F_{s}$ is the space time fermion number. 
The massless bosonic degrees of 
freedom in the NS-sector contain a sector identical with the type II 
theories but include the tachyon in addition. Furthermore 
the RR fields are doubled. Correspondingly there are two sorts of 
$Dp$ branes for $p$ odd. We denote them by $Dp_{\pm}$. The corresponding 
boundary states are \cite{Gaberdiel}
\beq
Dp_{\pm}\;\;:\qquad |Bp,\pm\>_{NSNS}+ |Bp,\pm\>_{RR},
\eeq
where $|Bp,\e\>_{\eta}$ satisfies 
\beqq
(\al_{n}^{\mu}-\tilde\al_{-n}^{\mu})|Bp,\e\>_{\eta}&=&0\n\\
(\psi_{r}^{\mu}-i\e\tilde\psi_{-r}^{\mu})|Bp,\e\>_{\eta}&=&0,
\eeqq
for $\mu\es 0,\cdots,p$ and 
\beqq
(\al_{n}^{\mu}+\tilde\al_{-n}^{\mu})|Bp,\e\>_{\eta}&=&0\n\\
(\psi_{r}^{\mu}+i\e\tilde\psi_{-r}^{\mu})|Bp,\e\>_{\eta}&=&0,
\eeqq
for $\mu\es p+1,\cdots,9$ and $\eta\in\{NSNS,RR\}$. 
Further information about the spectrum of type $O$B strings comes form 
the observation \cite{Gaberdiel} that only the combination 
$D_{p+}D_{p-}\equiv Dp_{\pm} $ which is charged only under the 
untwisted RR-fields has the correct number of degrees of freedom to 
be interpreted as the S-duals of the fundamental string and the 
NS-five brane respectively (the $D3_{\pm}$ is self dual 
\cite{Tseytlin}). The twisted 
fields should then be invariant under S-duality.  
\vspace{.3cm}
\pan
{\it Effective Action:}\pan
\pan
The string frame low energy effective action up to second order in the 
tachyon has 
been worked out in \cite{Tseytlin} (see also \cite{Grousi})  
\beqq\label{s10}
S&=&\frac{1}{2\kappa^{2} _{10}}\left[ \int \sqrt{-G}
\;e^{{-2\phi}}R+4(e^{{-2\phi}}\d\phi,\d\phi)-\frac{1}{4}(e^{{-2\phi}}\d T
,\d 
T)-\frac{m^{2}}{4}(e^{{-2\phi}}T,T)\right.\n\\
&&\left.-\frac{1}{2}(e^{{-2\phi}}H,H)
-\frac{1}{2}
(f_{+}(T)F_{[3]+},F_{[3]+})-\frac{1}{2}
(f_{-}(T)F_{[3]-},F_{[3]-})\right],
\eeqq
where $f_{\pm}(T)\es 1\pm T+\ha T^{2}+O(T^{2})$ and $(A,B)$ denotes 
the scalar product of differential forms. All the other 
fields are set to zero. It was argued in \cite{Tseytlin} that, 
provided the effective action is $S$-duality invariant, the exact form 
should be given by $f_{\pm}\es e^{\pm T}$. 

\section{Non-Dilatonic Solutions}
\def\p6{\phi_{6}}
All non-dilatonic solutions to the type 0 low energy effective 
action found so far had non-vanishing electric {\it and} magnetic 
RR-flux and correspond to appropriate projections of certain type II 
BPS solutions. However, as we shall argue below 
bound states that are purely electric or magnetic also exist 
but they cannot be obtained by as projections from the type 
II BPS states. 

We consider the six dimensional action obtained from \refs{s10} by 
parametrising $G_{10}$ 
in terms of the six dimensional Einstein metric
\beq
G_{10}=e^{\phi_{6}}g_{6}+e^{{2\nu}}dx_{I}dx^{I}.
\eeq
That is  
\beqq\label{e6}
S_{6}&=&\frac{1}{2\kappa^{2}_{6}}\left[ \int \sqrt{-G}R_{6} -
\frac{1}{2}(\d\phi_{6} ,\d\phi_{6} )-\frac{1}{4}(\d T
,\d T)-4(\d\nu,d\nu)-\frac{m^{2}}{4}(e^{{\phi}}T,T)\right.\n\\
&&\left.-\frac{1}{2}(e^{{-2\phi_{6} }}H,H)-
\frac{1}{2}(e^{4\nu} f_{+}(T)F_{[3]+},F_{[3]+})-\frac{1}{2}
(e^{4\nu} f_{-}(T)F_{[3]-},F_{[3]-})\right],
\eeqq
where $\phi_{6}\es \phi-2\nu$, $\kappa_{6}^{2}\es 
\kappa_{10}^{2}/L^{4}$ and we have performed the consistent 
truncation 
$B_{IJ}\es F_{I\mu\nu}\es H_{I\mu\nu}\es 0$. 
The relevant equations of motion that follow from \refs{e6} are given by 
\beqq
\nabla^2\phi_{6}&=&-\frac{m^{2}}{4}T^{2}e^{\p6}+(e^{-2\p6}H,H)\nn\\
\ha\nabla^2 T&=&-\frac{m^{2}}{2}Te^{\p6}-\ha(e^{4\nu} 
f_{+}'(T)F_{[3]+},F_{[3]+})-\frac{1}{2}
(e^{4\nu} f_{-}'(T)F_{[3]-},F_{[3]-})\n\\
8\nabla^2\nu&=&-2(e^{4\nu} f_{+}(T)F_{[3]+},F_{[3]+})-2
(e^{4\nu} f_{-}(T)F_{[3]-},F_{[3]-})\n\\
0&=&d*(He^{-2\p6})\\
0&=&d*(F_{[3]+}e^{4\nu}f_{+}(T)) \nn\\
0&=&d*(F_{[3]-}e^{4\nu}f_{-}(T)) \nn
\eeqq
supplemented with the charge quantisation conditions
\beq
Q_{1\pm}=\frac{1}{4\pi^{2}g\al'}\int\limits_{S^{3}}*e^{4\nu}F_{[3]\pm}\mtx{and} 
Q_{5\pm}=\frac{1}{4\pi^{2}g\al'}\int\limits_{S^{3}}F_{[3]\pm},
\eeq
and similarly for the NS field $H$. 
For radially symmetric configurations the latter can be implemented by 
writing the RR-$3$ form as ($r_{i}^{2} \es gQ_{i}\al'$)  
\beq\label{solveconstr}
F_{[3]\pm}=2r^{2}_{5\pm}\e_{3}+2r^{2}_{1\pm}e^{-4\nu}*e_{3},
\eeq
where $e_{3}$ is the volume form of the unit $3$-sphere. The 
`effective' potential for the tachyon on field and the size of 
the internal space $\nu$ is then found to be\footnote{To obtain the 
correct potential one proceeds, as in \cite{tseytlin0ld}, 
on the level of the equations of motion. Naive substitution of 
\refs{solveconstr} 
into the action leads to the wrong sign for the electric potential.} 
\beqq
V_{eff}(\nu,T)&=&\frac{m^{2}}{2}T^{2}e^{\p6}\\
&&+\frac{4}{V_{3}^{2}  }
\left(e^{-4\nu}(\frac{r^{4}_{1+}}{f_{+}(T)}+\frac{r^{4}_{1-}}{f_{-}(T)})
+e^{4\nu}(r^{4}_{5+}{f_{+}(T)}+r^{4}_{5-}{f_{-}(T)})\right),\n
\eeqq
where $V_{3}$ is the volume of the $3$-sphere. 
In the NS sector one finds a similar potential with $\nu$ 
replaced by $\p6$. 

Here we are interested in non-dilatonic solutions. Equating 
$\frac{\pa}{\pa T}V_{eff}\es \frac{\pa}{\pa \nu}V_{eff}\es 0$ we find 
\beqq\label{sys1}
T&=&0,\n\\
e^{8\nu}&=&\frac{r^{4}_{1+}+r^{4}_{1-}}{r^{4}_{5+}+r^{4}_{5-}}\\
0&=&e^{-4\nu}\left(-r_{1+}^4+r_{1-}^4\right)+e^{4\nu}
\left(r_{5+}^4-r_{5-}^4\right)\n
\eeqq
where we have used the fact that $f_{\pm}(0)\es\pm f_{\pm}'(0)\es 1$. 
Note that this is satisfied for any $RRT$ coupling $f_{\pm}(T)$ that 
agrees with $1\pm T$ up to $O(T)$. A particular solution of the the system 
\refs{sys1} is given by $r_{i+}\es r_{i-}, i\es 1,5$. This is the solution 
discussed in a paper by Costa \cite{Costa}. However, \refs{sys1} admits 
yet another set of solutions for $r_{i-}\es 0$ ($ r_{i+}\es 0$). 
This new class of black hole solutions has electric ($Dp_{+}$) brane charge 
but no magnetic charge. More precisely one finds  
\beq\label{e1}
e^{4\nu}=\frac{r^{2}_{1}}{r^{2}_{5}}\mtx{and}
ds_{6}^{2}=\frac{r^{2}}{r_{1}r_{5}}(-\d t^{2}+\d z^{2})+
\frac{r_{1}r_{5}}{r^{2}}(\d r^{2} +r^{2}\d\Omega_{3}^{2}).
\eeq
Note that we have chosen the solution which is $AdS_{2+1}$ for all $r$ 
(up to identifications). 
It corresponds to the near horizon regime of the 
asymptotically flat solution. The present solution is more suitable 
for reasons of stability but encodes the essential features of 
asymptotically flat black holes \cite{ivo}. 

The obvious question that arises now is whether the set of all these 
solutions is connected. This turns out to be the case. Indeed \refs{sys1} 
admits a family of solutions with 
\beq\label{inter}
r_{1+}=r_{5+}\equiv r_{+}\mtx{and}r_{1-}=r_{5-}\equiv r_{-}, 
\eeq
but $r_{+}\neq r_{-}$. This family interpolates between the set 
of $Dp_{\pm}$ black holes discussed by Costa \cite{Costa} and the 
the set of electric ($Dp_{+}$) solutions described above. The metric 
for the interpolating solutions \refs{inter} can be obtained from  
\refs{e1} upon substitution 
\beq
r_1^2=r_5^2 =\sqrt{r_+^4+r_-^4}.
\eeq
To summarise: \pan
{\it The low energy effective action for type $0B$ theory admits 
a continuous family of solutions interpolating between the 
$D1_{\pm}D5_{\pm}$ and the $D1_+D5_+$ configurations. This deformation 
preserves the non-dilatonic background.}\pan
\vspace{.1cm}

In the NS sector we parameterise $H$ by 
\beq
H=2r^{2}_{5}\e_{3}+2r^{2}_{1}e^{2\phi}*\e_{3}
\eeq
The corresponding solution is as in the type II case 
\beqq\label{NS}
ds_{6}^{2}&=&\frac{r^{2}}{r_{1}r_{5}}(-\d t^{2}+\d z^{2})+
\frac{r_{1}r_{5}}{r^{2}}(\d r^{2} +r^{2}\d\Omega_{3}^{2})\n\\
\nu&=& 0\\
e^{2\p6}&=&\frac{r^{2}_{5}}{r^{2}_{1}}.\n
\eeqq
Hence the metric is identical with the RR-solution. One might take 
this as an indication that that the configuration \refs{NS} is the $S$-dual 
of the $D1_{+}D5_{+}$ configuration. This is however misleading. As 
explained in section $2$ and \cite{Gaberdiel} the degrees of freedom 
of the $D1_{+}$ string do not match those of the fundamental string. 

To obtain the (extremal) black hole one proceeds as usual 
\cite{Gubser} by adding momentum in the $z$ direction, that is 
\beqq\label{bh}
ds_{6}^{2}&=&\frac{r^{2}}{r_{1}r_{5}}(-\d t^{2}+\d z^{2}+K(\d 
t+\d z)^{2} )+
\frac{r_{1}r_{5}}{r^{2}}(\d r^{2} +r^{2}\d\Omega_{3}^{2})\n\\
K&=& \frac{r_{K}^{2}}{r^{2}} 
\eeqq
The corresponding five-dimensional black hole is obtained by 
comparing \refs{bh} to the form
\beq
ds_{6}^{2}=e^{\frac{2}{3}\la}ds_{5}^{2} +e^{2\la}(\d z+A_{\mu}^{K})^{2}.
\eeq
The resulting (near horizon) five dimensional black hole 
metric is as in the type IIB 
case
\beqq
ds_{5}^{2}&=& -\frac{1}{f^{\frac{2}{3}}}\d t^{2}+f^{\frac{1}{3}}\left(\d 
r^{2}+r^{2}\d \Omega^{2}_{3}\right)\n\\
f&=&\frac{r^{2}_{1}r_{5}^2}{r^{4}}\left(1+\frac{r_K^2}{r^2}\right). 
\eeqq
\vfill\break
\pan
{\it Stability:}\pan
\pan
An important point in all type $0$ configurations is the stability. For 
the configuration with unequal electric- and magnetic charge the new 
issue arising is the mixing 
tachyonic- and $\nu$-modes\footnote{See also \cite{Myung} for an 
extensive discussion of stability for equal charges.} 
(recall $\nu$ describes the internal volume). 
More precisely the mass matrix becomes 
\beq\label{mass}
M=8\pmatrix{\frac{m^{2}}{8}+\frac{1}{r_{1} r_{5}}&& \frac{4q^2}{(r_{1} 
r_{5})^3}\cr \frac{4q^2}{(r_{1} r_{5})^3}&&\frac{16}{r_{1} r_{5}}},
\eeq
where $q\es r_{1+}^2r_{5+}^2-r_{1-}^2r_{5-}^2$ is the deformation parameter. 
On the other hand in the $6$-dimensional Einstein metric the radius 
of the $AdS_{3}$-factor is given by 
$l^{2}\es r_{1}r_{5}$. Linear stability for the scalar 
perturbations $T$ and $\nu$ then requires  
\beq
r_{1}r_{5}\leq-\frac{1}{\la_{-}},
\eeq
where $\la_{-}$ is the smaller eigenvalue of $M$. For $M$ as in \refs{mass} 
one then obtains 
\beq
-\frac{1}{4r_1r_5}\leq\left(\frac{17}{r_1 r_5}-\frac{1}{4\al'}\right)-
\sqrt{\left(\frac{15}{r_1 r_5}+\frac{1}{4\al'}\right)^2+
4\frac{16 q^2}{(r_{1} r_{5})^6}},
\eeq
where we have used that $m^{2}\es -2/\al'$. In the case of equal charges 
this leads to\footnote{The extra $\ha$ compared to \cite{Costa} comes when 
one takes into account that in $AdS_{2+1}$ stability is compatible with 
$m^2=-1/l^2$.} 
\beq
r_1r_5\leq (4+\ha)\al'.
\eeq
In the opposite limit, that is for $Q_{i-}\es 0$ one finds 
\beq
r_{1}r_{5}\leq\ha\left(\frac{137}{129}\right)\al'.
\eeq
In either case the 
curvature is of order of $\al'$ and hence corrections to the metric 
are to be expected unless the backgrounds is protected. It was observed 
before 
(e.g. \cite{Tseytlin,Costa}) that this applies to $AdS_{2+1}\times S^3$ 
backgrounds provided these 
corrections can be written in terms of the Weyl 
tensor only. 

For the F1NS5 configuration there is no mixing between the tachyon 
and $\nu$ and consequently linear stability requires 
\beq 
\frac{\al'}{2}\geq e^{\p6}r_{1}r_{5}=gQ_{5}\al',
\eeq
where $g$ is the string coupling. 

\section{World volume theory}
The metric of the $D1_{+}D5_{+}$ black hole described in the last 
section is identical with type IIB. The entropy and the 
Hawking decay rate are therefore also identical. If the black hole 
background described in the last section is a consistent 
ground state of string theory then it should be possible to 
reproduce these features form the world volume theory of 
the $D1_{+}D5_{+}$ system. This task is, however, complicated by 
the absence of supersymmetry or equivalently, by the presence of the 
tachyon which prevents us from interpolating the background to 
curvature much smaller than $1/\al'$. As a result the predictions  
for the world volume theory can be trusted only up to $\al'$- 
corrections. 

As a first step towards a detailed understanding of the world volume 
theory we discuss the existence an the nature of the $D1_{+}D5_{+}$ 
bound state\footnote{We refer the reader to \cite{Costa} for the analogous 
analysis of the $Dp_{\pm}$ bound states.}. 
For this we compute the open string loop amplitude in terms of the 
corresponding boundary states \cite{Gaberdiel}. For the NS-NS 
sector we have 
\beq
\<Bp,\e|e^{-lH_c^{osc}}|Bs,\e'\>_{{NSNS}}=\Tr_{NS}
\left(e^{-2lL_{0} }(\e\e')^{F}
\prod\limits_{\mu=p+3}^{s+2}(-1)^{B_{\mu} } \right),
\eeq
where 
\beq
(-1)^{B_{\mu} }\al_{m}^{\nu}(-1)^{B_{\mu} }=\cases{\al_{m}^{\nu}&
\hbox{$\mu\neq\nu$}\cr
-\al_{m}^{\nu}&\hbox{$\mu=\nu$}}
\eeq
and the same relation for the fermionic modes $\psi_{m}^{\nu}$. Here 
$H_c^{osc}$ is the oscillator part of the closed string Hamiltonian. 
The RR sector works analogously. 
The relevant amplitudes are then given by 
\beqq
\<Bp,+|e^{-tH_c^{osc}}|Bp,+\>&=& 
\frac{f_{3}^{8} (\tilde q)}{f_{1}^{8}(\tilde q)}-
\frac{f_{2}^{8}(\tilde q)}{f_{1}^{8}(\tilde q)}\mtx{and}\n\\
\<Bp,+|e^{-tH_c^{osc}}|Bs,+\>&=& 
\frac{f_{3}^{8-\nu}(\tilde q)}{f_{1}^{8-\nu}(\tilde q)}
\frac{f_{4}^{\nu} (\tilde q)}{f_{2}^{\nu} (\tilde q)}
\mtx{;}p<s,
\eeqq
where $\nu\es s-p$, $\tilde q\es e^{-2\pi l}$ and the functions 
$f_{i}(\tilde q)$ can be found in \cite{Cai}. In order to obtain the 
potential between two branes separated by a distance $r$, 
we perform the remaining integral over the momenta leading to 
\beqq
V_{p,p}(r)&=&-\int\frac{\d 
t}{2t}(8\pi^{2}\al't)^{-\frac{p+1}{2}}e^{{-t\tilde r^{2}}}
\left(\frac{f_{3}^{8} 
(q)}{f_{1}^{8}(q)}-\frac{f_{4}^{8}(q)}{f_{1}^{8}(q)}\right)\mtx{and}\n\\
V_{p,s}(r)&=&-\int\frac{\d 
t}{2t}(8\pi^{2}\al't)^{-\frac{p+1}{2}}e^{{-t\tilde r^{2}}}
\left(\frac{f_{3}^{8-\nu}(q)}{f_{1}^{8-\nu}(q)}\frac{f_{2}^{\nu} (q)}
{f_{4}^{\nu} (q)}\right),
\eeqq
where $\tilde r^{2} \es r^{2}/(2\pi\al')$, $q\es e^{-\pi t}$ and we 
have expressed the result 
in terms of the more familiar open string one loop amplitude via the 
substitution $l\es 1/(2t)$. At short distances which 
corresponds to $t\to \infty$, the potential between $D1_{+}$ and 
$D5_{+}$ branes is then obtained by expanding the integrand i.e. 
\beq\label{pot}
\frac{f_{3}^{8} (q)}{f_{1}^{8}(q)}-\frac{f_{4}^{8}(q)}{f_{1}^{8}(q)}= 
16+O(q)\mtx{and} 
\frac{f_{3}^{8-\nu}(q)}{f_{1}^{8-\nu}(q)}\frac{f_{2}^{\nu} (q)}
{f_{4}^{\nu} (q)}=2^{\nu/2} +O(q).
\eeq
As they stand the $t$-integrals are, however, not well defined due to 
the divergence at $t\to 0$. Of course, the full string theory result should 
be finite. As explained in \cite{TZ} the divergences in \refs{pot} 
are related to the presence of a tachyonic mode in the closed string sector 
and the coefficients of the counter terms needed to cancel the small $t$ 
divergences will depend on the 
details of the tachyon condensation. 
Fortunately for $V_{1,1}$ and $V_{1,5}$ it 
turns out that the counter term is $r$ independent and can 
therefore be discarded. For $r^{2}\!<\!<\!\al'$ and $\Lambda\es \sqrt{\al'}$ 
the resulting potentials read  
\beqq
V_{1,1}(r)&=& -\frac{\tilde r^{2}}{\pi^{2}\al'}\log(\tilde r^{2})\mtx{and}\n\\
V_{1,5}(r)&=& \frac{1}{4}V_{1,1}(r). 
\eeqq
In contrast to the $D3_{+}$ system \cite{Tseytlin}, the potential 
for $D1_{+}D1_{+}$ and for $D1_{+}D5_{+}$ is attractive for small 
distances. 
This is consistent with the existence of a conformal (non-dilatonic) 
gravitational background. 
For a $D5_{+}D5_{+}$-pair the potential takes  the 
form 
\beq
V_{5,5}(r)= 
c_{0}+c_{1}\tilde r^{2}
+c_{2}\tilde r^{4}+\frac{\tilde r^{6}}{3! 64\pi^{6}(\al')^{3}}
\log(\tilde r^{2}).
\eeq
As mentioned above the constants $c_{i}$ depend 
on the details of the tachyon condensation which is not understood in 
detail so far. We are therefore unable to fix them at present. The 
configuration with $\tilde r\es 0$ (i.e. all $D5_{+}$ on top of each 
other) is an extremum of $V_{5,5}(r)$ for 
all values of the $c_{i}$. However, the 
sign of $c_{1}$ determine whether this configuration is stable. If the 
stable configuration is at $r>0$ then, unless $Q_{5}\!<\!<\! Q_{1}$ 
most of the $(1,1)$ and the $(1,5)$ strings will be massive and will 
not contribute significantly to the entropy. If, on the other hand 
all of the $D5_{+}$ branes are on top of each 
other then the $(1,1)$ and the $(1,5)$ strings will be massless. 
However, a naive counting of the massless degrees of freedom 
still only accounts for $2/3$ of the Beckenstein-Hawking entropy. 
Indeed the $4Q_{1}Q_{5}$ bosonic degrees of freedom lead to a central 
charge \cite{Gubser}
\beq
c=4Q_{1}Q_{5}
\eeq
which is to be compared with $c\es 6Q_{1}Q_{5}$ in the type IIB 
case. 
The difference is, of course, due to the absence of fermions in the 
world volume theory. However, because the world volume theory is 
interacting even in the infrared the arguments presented above are 
probably too naive and a more quantitative analysis of the world volume 
theory will be needed.

\section{Discussion}
In this paper we have described a new class of non-dilatonic 
type $0$ backgrounds distinguished by the fact that only one of the 
two RR-fluxes ($F_{[n]\pm}$) is active. Unlike previous non-dilatonic 
backgrounds the solutions described here cannot be obtained by a 
projection of some type II BPS state. In particular we constructed 
an $AdS_{2+1}$ (BTZ) black hole from `electric' $D1_{+}D5_{+}$ branes. 
A novel feature of this configuration is the absence of fermions in 
the world volume theory. The existence of a non-dilatonic gravity dual 
suggests, however, that in the IR the world volume theory should nevertheless 
flow to a conformal field theory. In this paper we presented but a 
qualitative analysis the world volume theory. From a more detailed analysis 
of the interactions we may hope to gain valuable insight into the 
role of fermions in the microscopic theory of black holes. For that purpose 
the observation that the system can be continuously deformed form the 
$D1_{\pm}D5_{\pm}$ to the $D1_{+}D5_{+}$ configuration may prove useful. 
As 
all type $0$ configurations a tachyonic instability develops 
when the curvature decreases below the string scale. It seems 
reasonable to believe that this instability should have an analogue 
in the world volume theory, for example, in form of a large $N$ phase 
transition. A similar proposal has been made recently in 
\cite{Klebanov} for the world volume theory of the non-supersymmetric 
$D3$-brane system of type $0$ theory. It would be interesting to 
analyse the $D1D5$-brane system from this point of view.

\vspace{1cm}
\pan
{\bf Acknowledgements:}\pan
\vspace{.1cm}
\pan
The author would like to thank the Department of Mathematics at 
Kings College London for hospitality during the final stages of this 
work. This work was supported by a Swiss Government TMR Grant, BBW Nr. 
970557.

\end{document}